\documentclass[conference]{IEEEtran}
\IEEEoverridecommandlockouts

\usepackage{cite}
\usepackage{amsmath,amssymb,amsfonts}
\usepackage{algorithmic}
\usepackage{graphicx}
\usepackage{textcomp}
\def\BibTeX{{\rm B\kern-.05em{\sc i\kern-.025em b}\kern-.08em
    T\kern-.1667em\lower.7ex\hbox{E}\kern-.125emX}}

\usepackage{amsfonts}
\usepackage{amsmath}
\usepackage{booktabs}
\usepackage{array}
\usepackage{graphicx}
\usepackage{subcaption}
\usepackage{float}

\usepackage{bbm}
\usepackage{nicefrac}

\usepackage[utf8]{inputenc}
\usepackage{xspace}
\newcommand*{\eg}{e.g.\@\xspace}
\newcommand*{\ie}{i.e.\@\xspace}

\usepackage{ulem}
\usepackage[dvipsnames]{xcolor}

\usepackage{microtype}
\usepackage{graphicx}
\usepackage{booktabs} 

\usepackage{hyperref}

\begin{document}

\title{Revisiting Graph Projections for Effective Complementary Product Recommendation

}

\author{\IEEEauthorblockN{Leandro Anghinoni}
\IEEEauthorblockA{
\textit{MercadoLibre}\\
São Paulo, Brazil \\
leandro.anghinoni@mercadolivre.com}
\and
\IEEEauthorblockN{Pablo Zivic}
\IEEEauthorblockA{
\textit{MercadoLibre}\\
Buenos Aires, Argentina \\
pablo.zivic@mercadolibre.com}
\and
\IEEEauthorblockN{Jorge Adrian Sanchez}
\IEEEauthorblockA{
\textit{MercadoLibre}\\
Cordoba, Argentina \\
jorge.sanchez@mercadolibre.com}
}

\maketitle

\begin{abstract}
Complementary product recommendation is a powerful strategy to improve customer experience and retail sales. However, recommending the right product is not a simple task because of the noisy and sparse nature of user-item interactions. In this work, we propose a simple yet effective method to predict a list of complementary products given a query item, based on the structure of a directed weighted graph projected from the user-item bipartite graph. We revisit bipartite graph projections for recommender systems and propose a novel approach for inferring complementarity relationships from historical user-item interactions. We compare our model with recent methods from the literature and show, despite the simplicity of our approach, an average improvement of +43\% and +38\% over sequential and graph-based recommenders, respectively, over different benchmarks. 
\end{abstract}


\section{Introduction}
\label{introduction}

Complementary Product Recommendation (CPR) is an important area of research that significantly influences consumer behavior and business outcomes. CPR involves suggesting products that enhance or augment a customer's initial purchase, thereby creating value and improving the overall shopping experience \cite{li2024complementary}. For example, when a customer buys a cellphone, recommending complementary items such as a case or a charger increases the average transaction value and enhances customer satisfaction. The importance of complementary recommendations takes on special value in the industry and online commerce where the interplay between products can help uncover subtle purchase patterns, driving sales and encouraging loyalty.

The concept of complementarity in product recommendation refers to the degree to which two products can enhance each other's utility when consumed together. This concept translates into several different approaches in the literature, such as items bought together \cite{mcauley2015inferring}, purchase sequence \cite{yang2020large}, similarity in item content \cite{mcauley2015inferring, kang2019complete} or category relations \cite{yang2020large}. Although it is usual to assume the availability of a (partially) labeled dataset for training, building such a dataset is challenging not only due to the amount of data required to train today's most performant approaches but also due to the non-strict nature of the problem. Therefore, the standard approach is to infer complementarity relations from the users' interaction history. Collaborative filtering, content-based filtering, and hybrid models are commonly used techniques for analyzing large datasets of purchase histories and user interactions \cite{li2024complementary}. These algorithms extract patterns, identify relationships between products, and predict which items are likely to be purchased together.

Considering the temporal order of purchases is crucial for optimizing recommendations \cite{dang2023uniform}. When it comes to CPR, related products are usually close to each other in the purchase timeline of a user. Unlike personalization tasks, which require attending to the full history of user's interactions, complementarity is an item-item relation with a much shorter time span, and, therefore, methods that extract this relation explicitly may have an advantage over traditional sequential models. While personalization focuses on creating recommendations that align with the specific preferences and profiles of individual users, CPR operates on the principle of product relationships, emphasizing the synergy between products rather than the characteristics of the individual user. This means that while a personalized suggestion may arise from understanding the user's preferences, a complementary recommendation relies on identifying and suggesting products that are contextually linked to the item being considered.

Despite the recent advancements, several challenges can limit the performance of current CPR models. Graph neural networks, which have been the state-of-the-art (SOTA) for product recommendation, operate on simple graph structures, adopting simple rules for node connection, such as undirected user-item interactions. These types of connections do not consider global relations, which are hard to capture with traditional GNN frameworks. Learning strong item-item relationships can also be challenging for both sequential and GNN models, as the former operates on the whole user history and the latter on the whole set of user-item interactions at once. Training using negative samples can also become a limitation for large-scale problems, not to mention the lack of explainability of such methods.

Based on these observations, we revisit collaborative filtering via graph projection and add item-item temporal relations to build a meaningful and explainable graph structure that can be easily mined for product complementarity.

The contributions of this work are manyfold:

\begin{itemize}
    \item{The constructed graph is tailored for product complementarity, avoiding noisy relationships that may arise from considering the whole purchase history;}
    \item{The constructed graph is a one-mode projection over items, which captures better item-item complementarity;}
    \item{Inference is done by neighborhood search, in the form of graph adjacency, a local search that does not require calculating embedding similarity over large amounts of data;}
    \item{We propose an evaluation protocol, based on next item prediction, that makes our method comparable to sequential and GNN models.}
\end{itemize}

\begin{figure*}[htbp]
    \centering
    \begin{subfigure}[b]{0.3\textwidth}
        \centering
        \includegraphics[height=4cm, keepaspectratio]{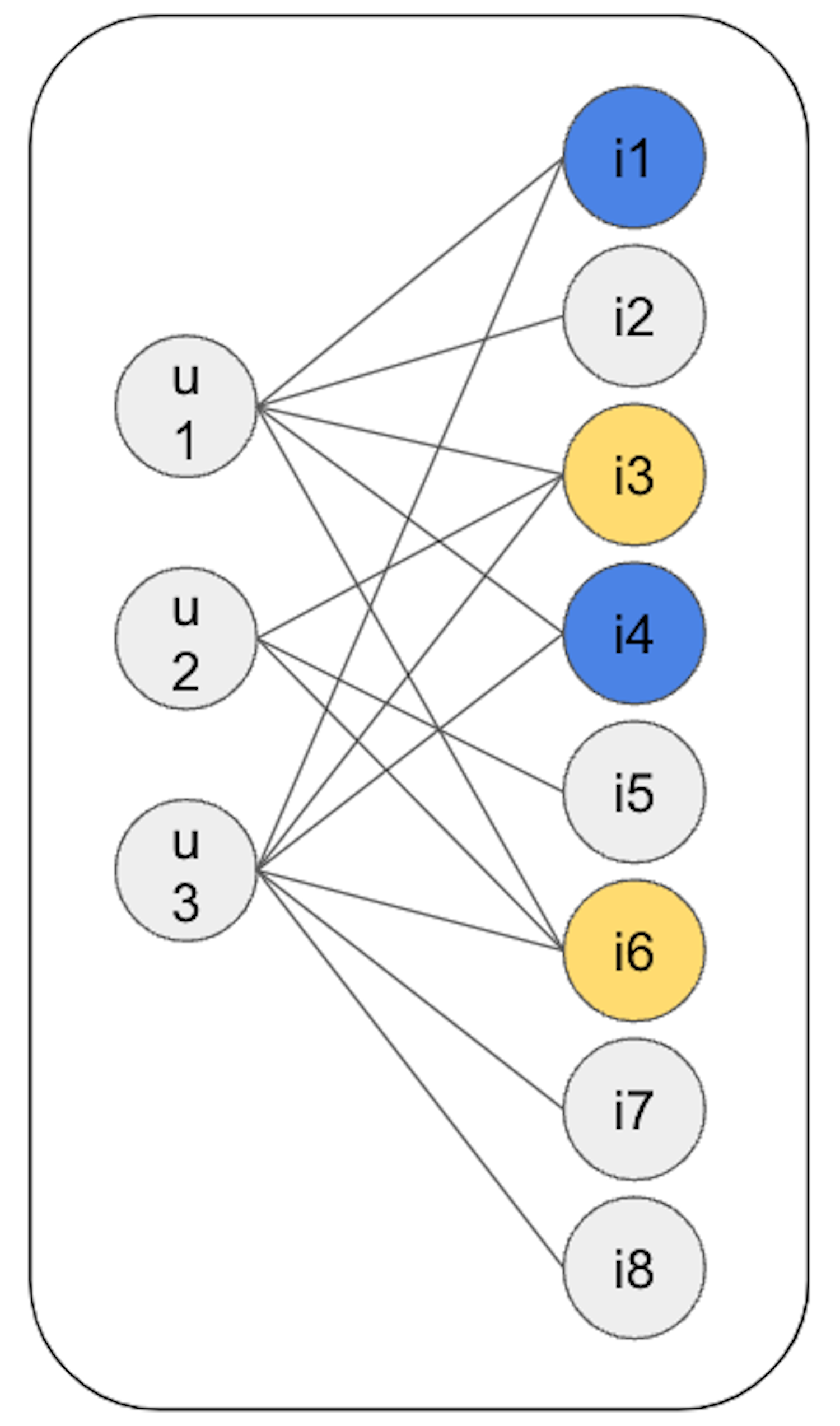}
        \caption{Bipartite graph}
    \end{subfigure}
    \hfill
    \begin{subfigure}[b]{0.3\textwidth}
        \centering
        \includegraphics[height=4cm, keepaspectratio]{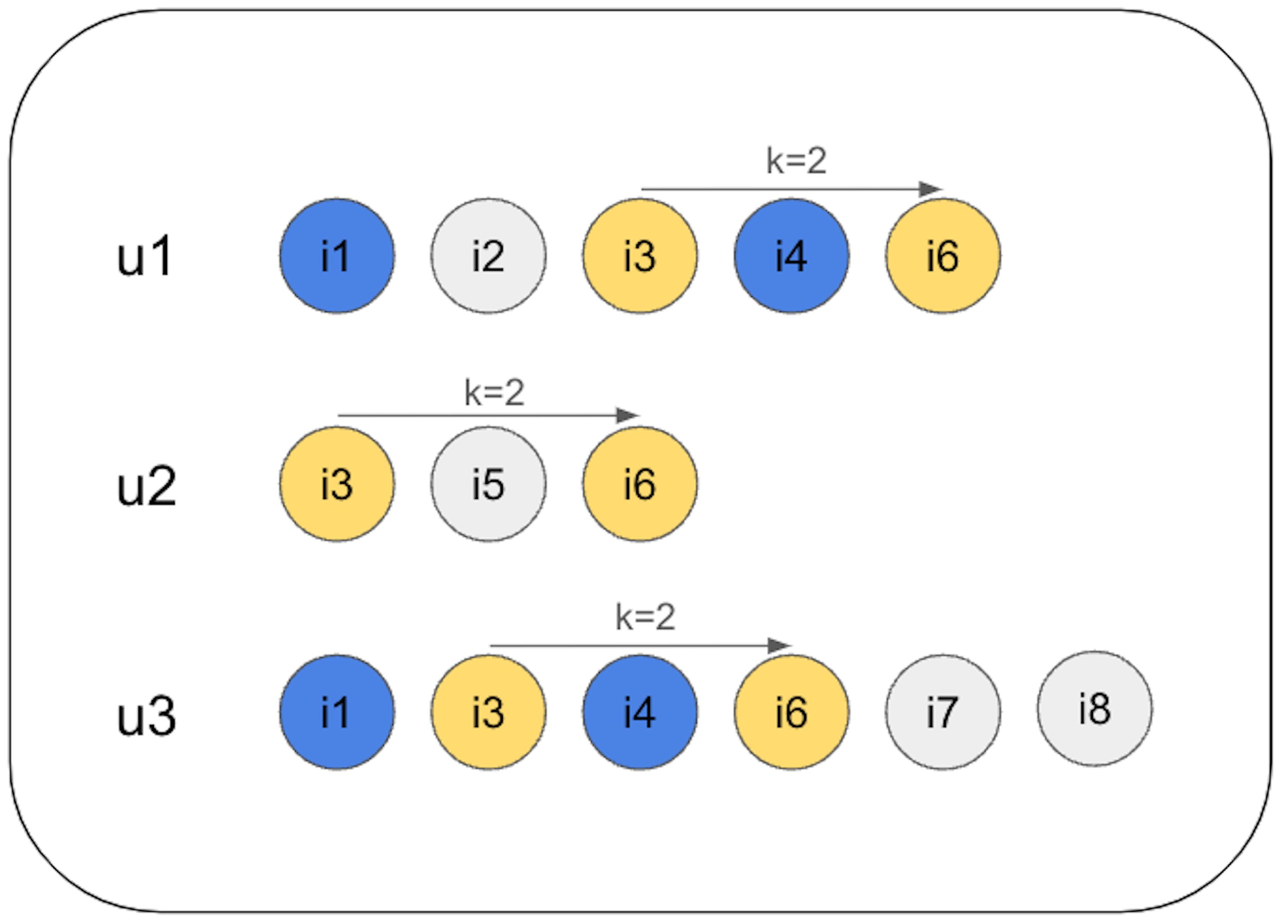}
        \caption{Purchase sequence}
    \end{subfigure}
    \hfill
    \begin{subfigure}[b]{0.3\textwidth}
        \centering
        \includegraphics[height=4cm, keepaspectratio]{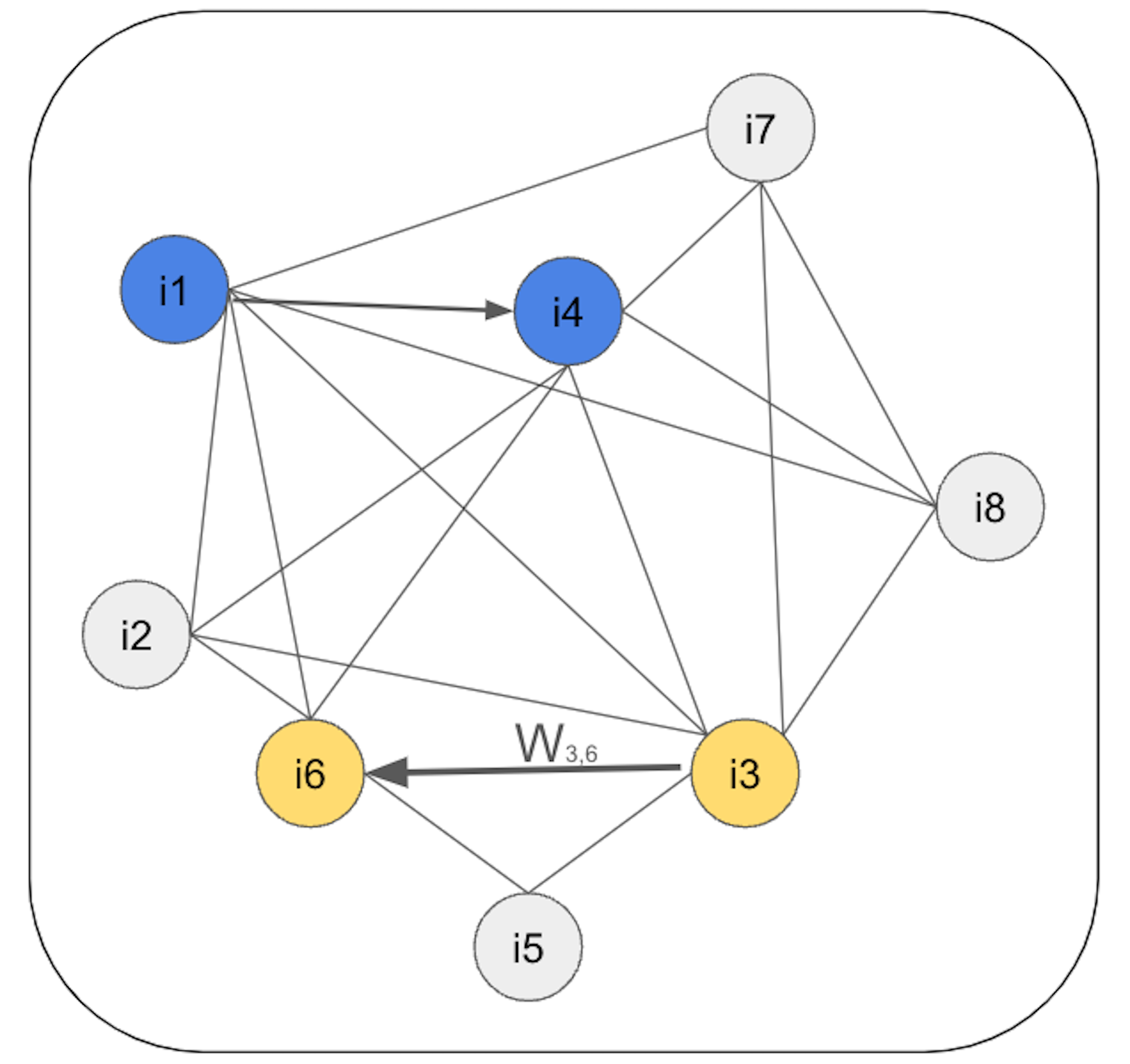}
        \caption{Final graph}
    \end{subfigure}
    \caption{Overview of the proposed model. In (a), items $i3$ and $i6$ should be strongly connected because they are bought together by all users. In (b) we set the direction of the complementarity by evaluating the purchase sequence of each user. In this case, $i6$ is always bought after $i3$. In (c) we construct the final graph considering the strengths of the connections in (a) and (b).}
    \label{fig:model}
\end{figure*}

\section{Related Work}
\label{related_work}

The definition of complementarity varies between studies. Several works \cite{yan2022personalized,mcauley2015inferring,liu2021item} define it based on the frequency of products being purchased together, while others \cite{zhang2021learning} consider products used together in specific scenarios. In the fashion domain, complementarity is also commonly defined based on the compatibility of clothing styles, materials, and sizes. Although it is also possible to frame the problem as a classification problem, using labeled data \cite{reddy2022shopping}, most works use user signal and product content to infer complementarity.

Current complementary recommendation models exhibit limitations in handling complex scenarios such as asymmetric complementarity, the coexistence of substitution and complementarity relationships, and the varying degrees of complementarity \cite{li2024complementary}. Addressing data sparsity, cold-start problems, and the need for explainability are also crucial for models in real-life application contexts. In what follows, we provide a brief overview of the literature grouped by topics most relevant to our work.

\paragraph{Collaborative Filtering via Graph Projection}

Since the seminal work of \cite{zhou2007bipartite}, several works have used the idea of an one-mode graph projection for personal recommendation \cite{shang2008personal,sawant2013collaborative,yang2020large}. Although promising, these early models were built to find user similarity and lack formality in terms of benchmark comparison. Inspired by the work of \cite{yang2020large}, which models both complementarity and supplementarity via graph projection, we revisit the topic by comparing our model to recent sequence-based models and neural graph models and by adopting metrics and data-sets that have become standard to product recommendation, such as NDCG@k and the Amazon Product Review Dataset \cite{mcauley2015image}.

\paragraph{Content-Based Models}

Content-based models leverage product attributes such as text descriptions and images to learn complementary relationships. Techniques such as Convolutional Neural Networks (CNNs) for image analysis and Natural Language Processing (NLP) for text analysis are commonly used \cite{mcauley2015inferring, kang2019complete}. However, these models often face challenges with data quality and the cold start problem for new products with limited information \cite{li2024complementary}.

\paragraph{Purchase Sequence-Based Models}

These models analyze user purchase sequences to identify complementary relationships. Methods like \mbox{\cite{wang2018path}} and \mbox{\cite{kvernadze2022two}} extract features using Item2Vec \cite{barkan2016item2vec} and Prod2Vec \cite{grbovic2015commerce} respectively to learn product embeddings based on co-occurrence patterns in user baskets. Although effective in capturing user-specific preferences, these models can be sensitive to noise in purchase data and may struggle to generalize to new or infrequent items.

Sequential recommenders, such as SASRec \cite{kang2018self} and BERT4Rec \cite{sun2019bert4rec}, learn user embeddings for personalized recommendation and have been a hot topic of research with a wide range of model variations to address different problems. They excel at recommending a personalized next product, however, when it comes to CPR, they may struggle due to the noise embedded in the whole user history.

\paragraph{Graph-Based Models}

Graph-based models represent products as nodes and relationships as edges in a graph. Graph Neural Networks (GNNs) are employed to learn product embeddings and predict complementary relationships \cite{zhang2021learning,hao2020p,liu2021item}. These models excel at capturing complex relationships but face challenges in constructing and maintaining accurate product relationship graphs.

Neural Graph Collaborative Filtering (NGCF) models integrate user-item interactions into the embedding process by utilizing a bipartite graph structure, allowing for the capture of high-order connectivity and collaborative signals that non-graph methods may overlook \cite{wang2019neural}. However, despite its strong representation capabilities, NGCF may still suffer from challenges related to model complexity and the need for careful hyper-parameter tuning to optimize performance across different datasets.

Tackling specific problems in CPR has also been the focus of several works, such as the cold-start and diversity problem. In \cite{hao2020p} the authors propose a method to provide diversified and relevant product recommendations, with good results in the cold-start scenario. They assume, however, a manual procedure for co-purchase data clean-up, that, although leads to better results, can be impractical in real e-commerce where products are constantly added to the dataset. 

Streamlined approaches have shown good results in collaborative filtering tasks. LightGCN \cite{he2020lightgcn}, for example, focuses solely on neighborhood aggregation by linearly propagating user and item embeddings through an interaction graph, using the weighted sum of embeddings from all layers to form the final representation. This eliminates unnecessary complexities like feature transformation and nonlinear activation, which have been found to contribute little to performance and complicate training. 

\paragraph{Recent trends}

Hybrid models combine multiple approaches, such as content-based and purchase sequence-based methods, to leverage their respective strengths \cite{li2024complementary}. While offering more robust recommendations, these models can be complex to implement and require careful tuning. Another recent approach is to use Large Language Models (LLM) to recommend items based on complementary concepts. In \cite{huang2023ccgen}, the authors propose a method to generate a list of complementary concepts along with explanations. The size of common complementary problems, however, is still a challenge to be processed through LLM calls.

\section{Proposed Method}
\label{sec:method}

In this Section, we propose a graph-based method that is simple, explainable, and aims to address the issues of noise and asymmetry mentioned above. We can break our method into two parts. First, similar to NGCF \cite{wang2019neural}, we define product complementarity by user co-purchase and project an item-item graph where the strength of the relations is defined by the probability that two items are connected through the same set of users. Second, we assign an order relation to the complementarity between any two items and adjust the strength of the connections based on the chronological span between these items in the user interaction history, limited to a time window of length $\kappa$.

The first step is very effective at linking complementary items and alleviates the problem of data noise by iterating over all users, \eg reducing the weight of items bought together only sporadically. The second part tackles the asymmetric nature of complementary relationships, by explicitly measuring the probability that an item is bought after the other within a certain period of time. Figure~\ref{fig:model} provides a schematic overview of each step.

Although our approach does not model cold-start scenarios explicitly, it can be easily adapted to deal with such situations. This will be discussed in Sec.~\ref{sec:cold-start}.

\subsection{Notation\label{sec:notation}}

Let $G = (U \cup V, \mathcal{E})$ be an unweighted, undirected bipartite graph, where $U$ and $V$ are disjoint sets of nodes, and $\mathcal{E} \subseteq \{(u, v) \mid u \in U, v \in V\}$ is the set of edges. We define the weighted projection graph onto $V$ as ${G}_V = (V, \mathcal{E}_V)$. An edge $(v_i, v_j)$ exists in $\mathcal{E}_V$ if and only if nodes $v_i$ and $v_j$ (where $i \neq j$) share at least one common neighbor in $U$ within the graph ${G}$. 

Each edge $(v_i, v_j) \in \mathcal{E}_V$ is associated with a positive weight $w_{ij} > 0$. These weights are symmetric, meaning $w_{ij} = w_{ji}$. The set of neighbors of a node $v_i \in V$ in the graph ${G}_V$ is denoted by $N(v_i)$ and defined as the set of vertices adjacent to $v_i$: $N(v_i) = \{v_j \in V \mid (v_i, v_j) \in \mathcal{E}_V\}$. The weighted degree of a node $v_i \in V$ in ${G}_V$, denoted $\text{deg}(v_i)$, is the sum of the weights of the edges incident to it, $\text{deg}(v_i) = \sum_{v_j \in N(v_i)} w_{ij}$.

\subsection{Problem Formulation}
Let $G$ be a bipartite graph, where the two disjoint sets of nodes $U$ and $V$ represent users and items respectively and $\mathcal{E}$ denote user-item the interactions between $(u_{i},v_{j})$. The goal is to find a graph representation $G_{V}$ of $G$ that retains items' relations found in the bipartite graph, \ie items that are bought together by different users should be strongly connected.

Item complementarity relationships are not necessarily symmetric, \ie item A can influence the purchase of B $(A \rightarrow B)$, but not necessarily the other way around. Take, for example, a cellphone and a cellphone case. A user will hardly buy the case before the phone. Other categories may have a more balanced relation, such as Fashion. As discussed in Sec.~\ref{introduction}, we use the temporal relationships between $(v_{i}, v_{j}) \in V$ in different timescales as a proxy to complementarity and incorporate this information to projection $G_{V}$. Assume items A and B are frequently bought by different users. Although the A and B are tightly connected in $G_{V}$ there is no direction in this connection. To infer item complementarity we add to this relation the number of times B is bought after A or vice-versa and consider the distance between user-item interactions. The underlying graph induced by these relations can be mined for complementarity by simply retrieving the set of neighbors $\mathcal{N}(v_{i})$ of any node $v_{i} \in V$ and ranking them by weights $w_{ij}$.

\subsection{One-mode Weighted Projection}

Inspired by \cite{zhou2007bipartite,yang2024efficient} we take a random walk approach to derive a projection of the bipartite graph $G$ over $V$.
Using the definitions introduced in Sec.~\ref{sec:notation}, we can write the following transition probabilities over $G$:
\begin{align}
p(u_i \to v_j) &= \frac{w_{ij}}{\text{deg}(u_i)} = \frac{w_{ij}}{\sum_{j} w_{ij}}, \label{eq:p_u_to_v} \\
p(v_j \to u_i) &= \frac{w_{ji}}{\text{deg}(v_j)} = \frac{w_{ji}}{\sum_{i} w_{ij}}\ . \label{eq:p_v_to_u}
\end{align}

Eq.~\eqref{eq:p_u_to_v} and \eqref{eq:p_v_to_u} encode one-step transition probabilities between nodes in $G$. Due to its bipartite nature, one-step transitions occur only between nodes from one set of nodes to the other. Let $P^{U\to V}$ the $|U|\times|V|$ matrix whose $ij$ entry equals $p_{u_i \to v_j}$, \ie the matrix of user-item normalized transition probabilities. Similarly, let $P^{V\to U}$ denote the $|V|\times|U|$ matrix corresponding to item-user transitions.

The probability of a walker traveling from node $v_i$ to node $v_j$ in exactly two steps is given by:
\begin{equation}
    p(v_i \to v_j) = p(v_i) \sum_{u_k \in U} p(v_i \to u_k) p(u_k \to v_j)\ ,\label{eq:p_vi_to_vj}
\end{equation}
where $p(v_i)$ denotes the probability that the walker starts at node $v_i$. We assume a uniform prior $p(v_i)=\nicefrac{1}{|V|}$, $\forall v_i\in V$ for simplicity. 
The probabilities in Eq.~\eqref{eq:p_vi_to_vj} can be written in matrix form as:
\begin{equation}
    Q = \frac{1}{|V|}\left(P^{V\to U} P^{U\to V}\right)
\end{equation}
In our case, we aim at capturing the probability of connecting two nodes, irrespective of the direction of the jumps ($v_i\to v_j$ or $v_j\to v_i$). In this case, such probabilities become:
\begin{equation}
    P^{V\leftrightarrow V} = Q + Q^T\ .
    \label{eq:p_v_v}
\end{equation}
We treat this $|V|\times |V|$ matrix $P^{V\leftrightarrow V}$ as the weighted adjacency matrix induced by an undirected weighted graph over $V$. Eq.~\eqref{eq:p_vi_to_vj} and \eqref{eq:p_v_v} resemble those from \cite{yang2024efficient}. Our formulation, however, avoids the inclusion of heuristics and additional transformations into the formulation.

We take this as the projection of $G$ over $V$ and define the strength of the connection between nodes in this graph as the weights obtained after a $\lambda$-step random walk on this graph, \ie
\begin{equation}
    W_V = {\left( P^{V\leftrightarrow V} \right)}^\lambda
    \label{eq:w_v}
\end{equation}

\subsection{Robust Estimation of Co-Purchase Directionality}
\label{sec:dwpg}

One of the characteristics of the graph induced by the projection onto $V$ is the symmetry in item interactions (see Eq.~\eqref{eq:p_v_v}). Although such behavior is desired from the affinity relations point of view, the temporal order in which two items are visited is highly relevant for complementarity. This Section introduces such additional constraints and proposes a final graph whose affinities can be mined for complementarity.

Let $C_I$ be a $|V|\times |V|$ matrix whose $ij$ entry encodes the number of times item $v_j$ was bought after item $v_i$ within a window of $\kappa$ steps. Entries of this matrix can be estimated simply by counting co-occurrences from historical data. In our case, we compute a variation of this idea that accounts for the greater relevance of those purchases that occur immediately after buying item $v_i$ \cite{dang2023uniform}, namely:
\begin{equation}
    {[C_I]}_{ij} = \sum_{n}{\mathbbm{1}[\delta^{i\to j}_n \leq \kappa]\left(\frac{1}{\delta^{i\to j}_n}\right)}\ .
    \label{eq:cij}
\end{equation}
$\mathbbm{1}[\cdot]$ is the indicator function, and $\delta^{i\to j}$ denotes the number of steps a user spent in buying item $v_j$ after having bought item $v_i$. Note that $\delta^{i\to j}$ accounts for interaction steps instead of the absolute time between purchase events. The choice of aggregation function is further discussed in Sec.~\ref{sec:ablation-k}.

Since Eq.~\eqref{eq:cij} is based on co-occurrence statistics, it suffers from the sparsity induced by large product catalogs. To account for this problem, we define an auxiliary $|V|\times |V|$ matrix $C_C$ as before but, instead of counting co-occurrences at the item level, we look at the categories of item pairs. The rationale is the same as that of Eq.~\eqref{eq:cij} but looking at the number of times a category $y_j$ was bought after buying an item of category $y_i$. This coarsening of the co-counts stored in $C_C$ acts as a regularizer that reduces the sparsity in the co-counts. We denote by $\tilde{C}_I$ and $\tilde{C}_C$ the row-normalized versions of $C_I$ and $C_C$, respectively.

Finally, we define $W$ as the Hadamard product between $W_{V}$ (Eq.~\eqref{eq:w_v}) and the convex combinations of $\tilde{C}_I$ and $\tilde{C}_C$:
\begin{equation}
    W = W_{V} \odot \left((1-\alpha)\tilde{C}_I+\alpha\tilde{C}_C\right)\ .
    \label{eq:W}
\end{equation}

Eq.~\eqref{eq:W} can be seen as the adjacency matrix of a graph over $V$ with the property that for every item $v_{i}$, there is a set of neighbors $N(v_{i})$ which are defined by the co-purchase graph projection and directed and weighted by complementarity relation in user's history. To retrieve a set of complementary items of $v_{i}$ we rank all edges $(v_i, v_j)$ by their weights, \ie by looking at the off-diagonal coefficients of the $i$-th row of $W$.

\subsection{Time complexity}

The overall complexity of the proposed model is $\mathcal{O}(\log \lambda N_{nz})$, where $\lambda$ is the exponent and $N_{nz}$ is the number of non-zero entries in Eq. (\ref{eq:w_v}). This assumes that $W$ remains sparse after $\lambda$ exponentiation, and we use exponentiation by squaring. All other data structures can be built, in the worst case, with complexity linear in the number of user-item iterations $\mathcal{O}(N)$.

In practical applications, scalability can be improved using simple heuristics. For instance, we can split the graph into subgraphs and process them in parallel. Also, the parameter $\lambda$ is tuned according to graph size and sparsity, with lower values proving effective for larger, denser graphs.

\section{Experiments}
\label{sec: experiments}

\begin{table*}[ht]
    \centering
    \caption{Performance comparison on sequential recommendation. *We follow the same setup as in \cite{rajput2024recommender}.
    }
    \resizebox{\textwidth}{!}
    {
    \begin{tabular}{@{}>{\raggedright}m{1.8cm} c c c c c c c c c c c c@{}}
        \toprule
        \textbf{Methods} & \multicolumn{4}{c}{\textbf{Sports and Outdoors}} & \multicolumn{4}{c}{\textbf{Beauty}} & \multicolumn{4}{c}{\textbf{Toys and Games}} \\ \cmidrule(lr){2-5} \cmidrule(lr){6-9} \cmidrule(lr){10-13}
        & \textbf{Rec@5} & \textbf{ndcg@5} & \textbf{Rec@10} & \textbf{ndcg@10} & \textbf{Rec@5} & \textbf{ndcg@5} & \textbf{Rec@10} & \textbf{ndcg@10} & \textbf{Rec@5} & \textbf{ndcg@5} & \textbf{Rec@10} & \textbf{ndcg@10} \\ 
        \midrule
        P5\cite{geng2022recommendation}       & 0.0061 & 0.0041 & 0.0095 & 0.0052 & 0.0163 & 0.0107 & 0.0254 & 0.0136 & 0.0070 & 0.0050 & 0.0121 & 0.0066\\ 
        Caser\cite{tang2018personalized}    & 0.0116 & 0.0072 & 0.0194 & 0.0097 & 0.0205 & 0.0131 & 0.0347 & 0.0176& 0.0166 & 0.0107 & 0.0270 & 0.0141\\ 
        HGN\cite{ma2019hierarchical}      & 0.0189 & 0.0120 & 0.0313 & 0.0159 & 0.0325 & 0.0206 & 0.0512 & 0.0266 & 0.0321 & 0.0221 & 0.0497 & 0.0277\\ 
        GRU4Rec\cite{hidasi2015session}  & 0.0129 & 0.0086 & 0.0204 & 0.0110 & 0.0164 & 0.0099 & 0.0283 & 0.0137 & 0.0097 & 0.0059 & 0.0176 & 0.0084\\ 
        BERT4Rec\cite{sun2019bert4rec} & 0.0115 & 0.0075 & 0.0191 & 0.0099 & 0.0203 & 0.0124 & 0.0347 & 0.0170 & 0.0116 & 0.0071 & 0.0203 & 0.0099\\ 
        FDSA\cite{zhang2019feature}     & 0.0182 & 0.0122 & 0.0288 & 0.0156 & 0.0267 & 0.0163 & 0.0407 & 0.0208 & 0.0228 & 0.0140 & 0.0381 & 0.0189\\ 
        SASRec\cite{kang2018self}   & 0.0233 & 0.0154 & 0.0350 & 0.0192 & 0.0387 & 0.0249 & 0.0605 & 0.0318 & 0.0463 & 0.0306 & 0.0675 & 0.0374 \\ 
        S$^{3}$-Rec\cite{zhou2020s3}  & 0.0251 & 0.0161 & 0.0385 & 0.0204 & 0.0387 & 0.0244 & 0.0647 & 0.0327 & 0.0443 & 0.0294 & 0.0700 & 0.0376\\ 
        TIGER*\cite{rajput2024recommender}       & \underline{0.0264} & \underline{0.0181} & \underline{0.0400} & \underline{0.0225} & \underline{0.0454} & \underline{0.0321} & \underline{0.0648} & \underline{0.0384} & \underline{0.0521} & \underline{0.0371} & \underline{0.0712} & \underline{0.0432}\\
        \midrule
        Ours  & \textbf{0.0360} & \textbf{0.0260} & \textbf{0.0480} & \textbf{0.0299} & \textbf{0.0669} & \textbf{0.0488} & \textbf{0.0851} & \textbf{0.0546} & \textbf{0.0799} & \textbf{0.0598} & \textbf{0.1010} & \textbf{0.0666}\\ 
        & +36.4\%&+43.6\%&+20.0\%&+32.9\%&+47.3\%&+52.0\%&+31.3\%&+42.2\%&+53.3\%&+61.2\%&+41.8\%&+54.2\%\\ 
        \bottomrule
    \end{tabular}
    }
    \label{tab:performance_comparison_sequential}
\end{table*}

\subsection{Datasets}

We use four different subsets from the Amazon Product Reviews Dataset \cite{mcauley2015image}, namely: "Beauty", "Sports and Outdoors" (\textit{Sports} for short), "Toys and Games" (\textit{Toys} for short), and "Books". These subsets differ not only in the number of active users and items but also in the number of interactions among them. Table~\ref{tab:data_stats} shows summary statistics for each of these sets. For all datasets, we define a complementary item as an item bought together or after the query item. The datasets are filtered using a $n$-core procedure, i.e., only users and items with at least $n$ interactions are kept in the dataset. For sequential recommendation, we use a 5-core setup, and for graph-based recommendations, we use a 10-core setup as reported in \cite{rajput2024recommender} and \cite{wang2019neural}, respectively.

\begin{table}[ht]
    \centering
    \caption{Statistics for datasets used in this work}
    {
    \begin{tabular}{@{}lccc@{}}
        \toprule
        Dataset & \# Users & \# Items & \# Interactions \\ \midrule
        Beauty & 22,363 & 12,101 &  198,502\\
        Sports and Outdoors & 35,598 & 18,357 &  296,337\\
        Toys and Games & 19,412 & 11,924 & 167,597 \\ 
        Books & 52,643 & 91,599 & 2,984,108\\
        \bottomrule
    \end{tabular}
    }
    \label{tab:data_stats}
\end{table}

\subsection{Experimental Setup}

All datasets are ordered by user ID and by timestamp. We adopt different split procedures for sequential and graph-based models to match the experimental setup of others' works. Models parameters $(\alpha, \lambda, \kappa)$ are optimized in a validation split. All experiments use Recall@k and NDCG@k as metrics.

\subsection{Sequential Recommendation}

The splitting strategy for sequential recommendation models is as follows: the last item of each interaction sequence is used for testing purposes, the second last for validation, and the rest is used for training \cite{rajput2024recommender}. We perform a grid search over the hyperparameter space by using the training and validation data to build the matrices in Eq.~\eqref{eq:W} and to evaluate model performance, respectively. For evaluation, we take the last item of each training sequence as the query item and compare the model prediction against the validation sample. Once we identified the optimal hyperparameter combination, we use these values to re-compute the matrices using both training and validation data. We then compute the final performance on the test set.
By this procedure, the optimal parameter triplet $(\alpha, \lambda, \kappa)$ for Sports, Beauty, and Toys are $(0.85, 4, 2)$, $(1, 4, 4)$, and $(0.9, 4, 4)$, respectively. From these numbers we can infer the following: (i) in our model, item-item relation is more important than category-category relation, (ii) aggregating information from too many steps away is not beneficial (this is further explored in Sec.~\ref{sec: ablation-lambda}) and (iii) our model captures short-term complementarity, no longer than $\kappa$ purchases away (see more in Sec.~\ref{sec:ablation-k}).

Results for complementary item recommendation using sequential models are shown in Table~\ref{tab:performance_comparison_sequential}. In the table, we show Recall@k and NDCG@k (k=5 and 10) for different models proposed in the literature as well as for our approach. From the table, we see that our model outperforms sequential recommenders by a large margin (+43\% on average) in all datasets and all metrics.

A key difference between our method and those reported in Table~\ref{tab:performance_comparison_sequential} is that, although we consider the whole user interaction, we explicitly model item-item relation within a window of $\kappa$ steps. Models such as SASRec \cite{kang2018self} and BERT4Rec \cite{sun2019bert4rec} learn a high dimensional embedding for each item and predict the next item by approximate nearest neighbor search. The previous best-performing method \cite{rajput2024recommender} uses a transformer-based encoder-decoder setup and is trained in a sequence-to-sequence fashion.

\subsection{Graph-based Recommendation}

We also compare our approach with graph-based models. In this case, we follow different splitting and evaluation strategies to match the common practice in the literature \mbox{\cite{wang2019neural}}. We take a random 80\% of its interactions for each user for training. From the remaining 20\%, we take one half for validation and the other half for testing, respectively. Note that, different from the sequential case, each user might have more than a single target. For evaluation, we take as a query the item that precedes the test (or validation) sample. During the construction of the matrices in Eq.~\eqref{eq:W}, we ignore all connections that start or end in a test item.
The final result is reported using ($\alpha$, $\lambda$, $\kappa$) = (1, 1, 4). Table~\ref{tab:performance_comparison_graphs} shows results for a variety of graph-based methods from the literature. As in the case of sequential recommendations, our approach outperforms them by a considerable margin despite its greater simplicity.

A common characteristic of the best performing methods is the simplification of GCN-based collaborative filtering models, which are hard to train and adopt in industry \cite{mao2021ultragcn}. This paradigm change was also present in \cite{he2020lightgcn} and indicates that simpler models not only can lead to better results but also training efficiency.

\begin{table}[ht]
    \centering
    \caption{Performance comparison on graph-based recommendation.}
    {
    \begin{tabular}{@{}>{\raggedright}m{3cm} c c @{}}
        \toprule
        \textbf{Methods} & \multicolumn{2}{c}{\textbf{Books}} \\ 
        \cmidrule(lr){2-3} 
        & \textbf{Recall@20} & \textbf{ndcg@20} \\ 
        \midrule
        MF-BPR\cite{koren2009matrix}	  &0.0338	&0.0261\\
        CML\cite{hsieh2017collaborative}	      &0.0522	&0.0428\\
        ENMF\cite{chen2020efficient}	  &0.0359&	0.0281\\
        DeepWalk\cite{perozzi2014deepwalk}  &0.0346&	0.0264\\
        LINE\cite{tang2015line}	  &0.0410&	0.0318\\
        Node2Vec\cite{grover2016node2vec}  &0.0402&	0.0309\\
        NGCF\cite{wang2019neural}	 &0.0344	&0.0263\\
        NIA-GCN\cite{sun2020neighbor}	 &0.0369	&0.0287\\
        LR-GCCF\cite{chen2020revisiting}	 &0.0335&	0.0265\\
        LightGCN\cite{he2020lightgcn} &0.0411	&0.0315\\
        DGCF\cite{wang2020disentangled}	 &0.0422	&0.0324\\
        UltraGCN{base}	&0.0504&	0.0393\\
        UltraGCN\cite{mao2021ultragcn} &\underline{0.0681}	&\underline{0.0556}\\
        \midrule 
        Ours   &  \textbf{0.1053} &  \textbf{0.0702} \\ 
        & +54.6\% & +26.3\% \\ 
        \bottomrule
    \end{tabular}
    }
    \label{tab:performance_comparison_graphs}
\end{table}

\subsection{Complementarity and the problem of cold-starts.}
\label{sec:cold-start}

Benchmark setups filter user and item interactions (keeping 5-core or 10-core interactions) to control sparsity and be representative of a wide variety of models. Although useful from a benchmarking point of view, this configuration hides the cold-start phenomena faced by most practical applications. Although not in the main scope of our work, we illustrate a simple strategy to deal with this problem. In this exercise, a cold-start item is an item with no history in users' interactions, but for which we would like to generate a complementary recommendation anyway. Table~\ref{tab:amazon_toys_ndcg} shows results for different cold-start rates on the Toys dataset. Performance is disaggregated for both warm and cold items. We show two different scenarios: the raw application of our approach for both cold and warm items (noted as 'Ours' in Table. \ref{tab:amazon_toys_ndcg}), and the extension of our model by a simple strategy based on similarities between cold and previously seen items (noted as 'Ours + affinity' in Table. \ref{tab:amazon_toys_ndcg}). Details on the specific procedure are given next.

First, we split the dataset into train and test sets with a ratio of 90/10. Next, we randomly choose 2, 5, and 10\% of items and remove them from the training data split, similar to \mbox{\cite{rajput2024recommender}}. During evaluation, if a test item belongs to the cold-start group, we search for the most similar item in the other group. By doing so, we leverage the relations learned for a similar item in terms of co-purchase strength and directionality. We do so by first restricting the search space to items in the same sub-category as the cold-start query, and looking at the most similar item within this set. We use the cosine similarity between title embeddings as a proxy for item affinity. To compute title embeddings, we use the \texttt{all-MiniLM-L6-v2} model from SBERT \cite{reimers-2019-sentence-bert}. Current solutions to the cold-start problem include transfer learning, considering contextual product information and generative models \cite{li2024complementary}. They share the idea of finding a good representation for the cold item (or user) in the projected space, which we adopted in a simpler way.

As we see from the table, performance degrades as the percentage of cold-start items increases, which is to be expected. We also see that when the basic model (Ours) is faced with cold-start items, performance is zero as the model cannot recognize them as part of the underlying graph. Only when we add some heuristics based on affinities between cold and warm items (Ours + affinity), the model is able to cast recommendations for these items too. Performance in this case depends heavily on the ability of the model to relate cold items to warm-start items. In our case, the approach we introduced above provides a simple baseline that can serve as a reference for future evaluations.

\begin{table}[ht]
    \centering
    \caption{Item cold-Start on the Amazon Toys dataset. Performance is measured by NDCG@5.}
    {
    \begin{tabular}{@{}lccc@{}}
        \toprule
        \textbf{Cold-start rate} & 2\% & 5\% & 10\%  \\ \midrule
        Ours & 0.0485 & 0.0403 & 0.0335  \\
        \ \ \ warm items & 0.0555 & 0.0564 & 0.0612  \\
        \ \ \ cold items & 0.0000 & 0.0000 & 0.0000  \\ \midrule
        Ours + affinity & 0.0502 & 0.0432 & 0.0368  \\
        \ \ \ warm items & 0.0555 & 0.0564 & 0.0612  \\
        \ \ \ cold items & 0.0133 & 0.0102 & 0.0073  \\ 
        \bottomrule
    \end{tabular}
    }
    \label{tab:amazon_toys_ndcg}
\end{table}

\subsection{Ablations}
\label{sec:ablation}
In this section, we present a series of ablations to our framework to identify which components play a key role.

\begin{figure*}[htbp]
    \centering
    \begin{subfigure}[b]{0.3\linewidth}
        \centering
        \includegraphics[width=1\linewidth]{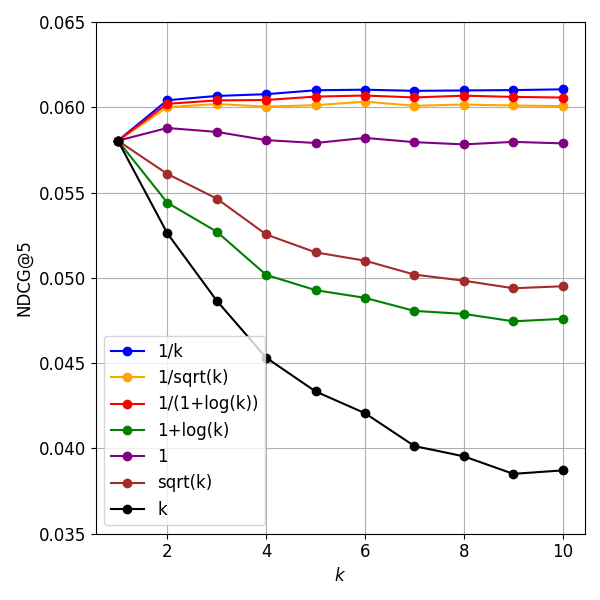}
        \caption{Temporal aggregation}
        \label{fig:ablation_k}
    \end{subfigure}
    \hfill
    \begin{subfigure}[b]{0.3\linewidth}
        \centering
        \includegraphics[width=1\linewidth]{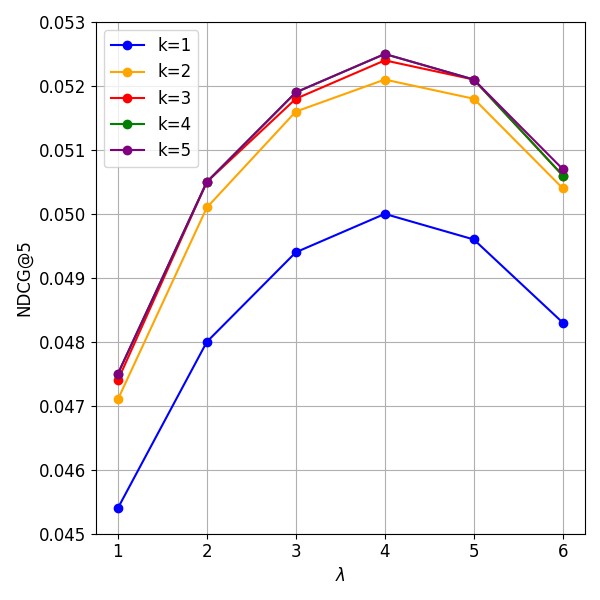}
        \caption{High-order relations}
        \label{fig:ablation_lambda}
    \end{subfigure}
    \hfill
    \begin{subfigure}[b]{0.3\linewidth}
        \centering
        \includegraphics[width=1\linewidth]{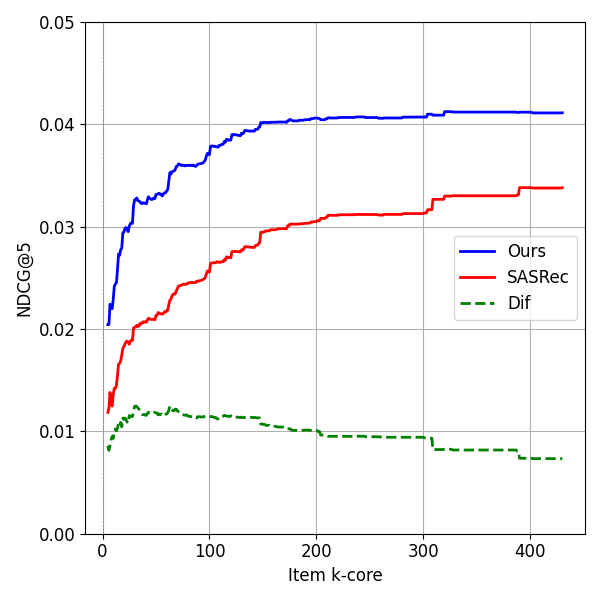}
        \caption{Test item k-core}
        \label{fig:ablation_kcore}
    \end{subfigure}
    \caption{Ablation studies on Amazon Beauty. In all studies we evaluate NDCG@5. In (a) we show the effect of different temporal aggregation functions. Giving less weight to purchases further away is beneficial for CPR. In (b) we see that there is an optimum point for high-order aggregation, \ie , aggregating information from users too many hops away can harm CPR performance. Finally, in (c) we show that our model shows a higher performance over the full k-core range, reaching the asymptotic performance values earlier than SASRec.}
    \label{fig:ndcg_comparison}
\end{figure*}

\subsubsection{Time-continuous vs. step-wise encoding}
\label{sec:ablation-k}

As described in Sec.~\ref{sec:method}, our method attributes a certain transition probability from a query item to a complementary item based on the temporal distance of the two purchases. There are two aspects to be considered in this approach: (i) if the actual time between interactions should be considered and (ii) how to differentiate purchases that are closer in time from those that are distant. As for (i), we experimented with both the actual time difference between purchases, $k=|t_j-t_i|$ \footnote{$t_i$ denotes the Linux time stamp (measured in days) of the purchase of $v_i$.}, and step-wise differences, $k\in(1,\kappa)$, as presented in Sec.~\ref{sec:dwpg}. In preliminary experiments, we observed that both approaches obtained similar performances. We opted for the step-wise approach for its simplicity and previous works \cite{dang2023uniform}. 

As for (ii), we evaluated seven different aggregation functions to weigh the contribution of the items in the user interaction history, these functions give different weights to the events in time, but those giving higher weights to closer-to-the-query events show better performance, as shown in Fig.~\ref{fig:ablation_k}.

\subsubsection{Effects of High-order Relations}
\label{sec: ablation-lambda}

Parameter $\lambda$ in Eq.~\eqref{eq:w_v} can be seen as the number of steps incurred by a random walker in the projection graph associated with the adjacency matrix in Eq.~\eqref{eq:p_v_v}. A higher value of $\lambda$ relates to the number of different ways two nodes (items) can be connected by a walk. Our results are consistent with those from \mbox{\cite{he2020lightgcn}}, \ie there is an optimal value for aggregating these higher-order relations. In our case, that is tuned by $\lambda$, in \cite{he2020lightgcn} that is tuned by the number of layers of the GCN. Figure \ref{fig:ablation_lambda} shows the optimal $\lambda$ for the Amazon Beauty dataset. We fix $\alpha=1$ and plot the curves for different values of $k$, finding the optimal value $\lambda = 4$.

\subsubsection{On Data Sparsity}
Data sparsity in the context of recommender systems refers to the effects produced by low-count user-item co-occurrences. Common benchmarks overcome this difficulty by imposing strict filtering conditions, requiring a minimum count of interaction both at the item and user level, \ie k-core sets. Fig.~\ref{fig:ablation_kcore} illustrates the performance of our model as well as that of SASRec for increasing values of k, for the Beauty subset of the Amazon Products Review dataset. We observe similar dynamics for both methods. However, our model shows a higher performance over the full k-core range, reaching the asymptotic performance values earlier than SASRec.

\subsubsection{On Data Noise}
\label{sec:noise}
\begin{figure}[ht]
    \centering
    \includegraphics[width=0.9\columnwidth]{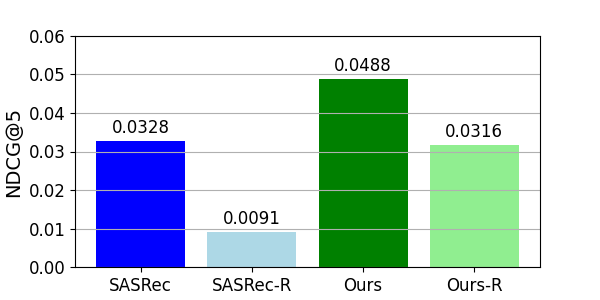} 
    \caption{Performance under permutation of items with the same timestamp}
    \label{fig:img_noise}
\end{figure}

Noise is another challenge for recommender systems, specially to models that infer complementary relationship from co-purchase data. Complementarity can be weakly directed (any product can come first) and users may perform other purchases in between. Another issue with a more practical application is items bought together. These items share the same timestamp and add a bias related to the original order of these items in the data. We found that, for the Amazon Beauty dataset, 39\% of the purchases share the same timestamp with the consecutive purchase (the purchases are ordered by \texttt{asin} by default). We removed this bias by randomly sorting these bought-together items. Figure~\ref{fig:img_noise} shows NDCG@5 for the original dataset and the mean of a 5-run for the shuffled dataset (-R). Our model presents a much lower degradation than SASRec (-35\% vs. -72\%), which is more sensitive to sequence ordering due to the presence of positional embeddings, especially in sparse datasets \cite{kang2018self}.

\subsection{Qualitative examples}
\label{sec:quali_ex}
\begin{figure*}[ht]
    \centering
    \includegraphics[width=0.95\linewidth]{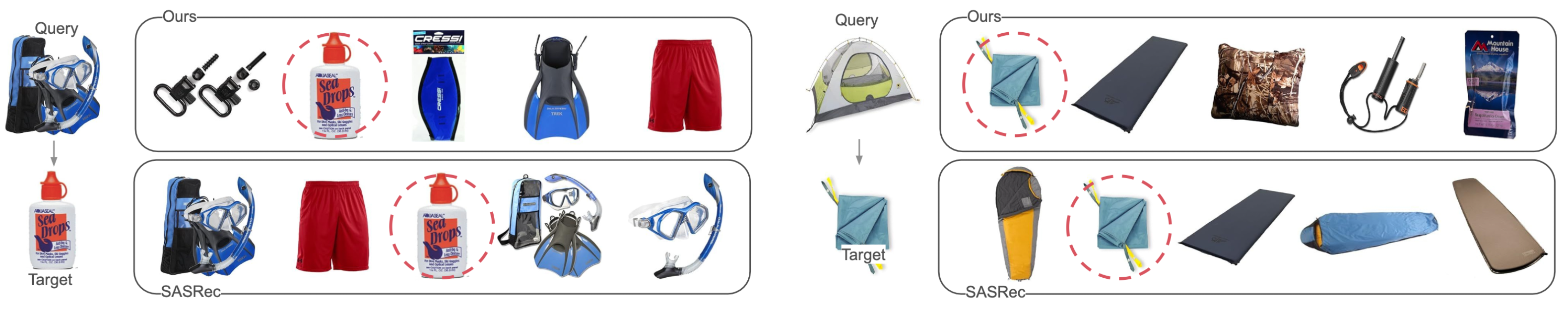}
    \caption{Qualitative examples for Our model (on top) vs. SASRec (below). Notice that our model not only ranks higher than the test item but also includes only complementary items in the top-k items. SASRec recommendations lack diversity.}
    \label{fig:cpr_example_}
\end{figure*}

We depict in Fig.~\ref{fig:cpr_example_} two examples of the complementary items generated by Our model and SASRec. In both cases, Our model ranks higher than the test item. Besides that, the products generated by Our model do not contain items similar to the query item (as the snorkeling set in the first example) and present a wider variety of complementary items. Notice that, in the second example, SASRec recommends several similar sleeping bags.

\section{Conclusion}
\label{sec:conclusion}

In this work, we introduced a model for recommending complementary products based on a query item. Despite its simplicity, our model significantly outperforms the benchmarks considered. By explicitly incorporating strong item-item relationships into the graph through collaborative filtering and the temporal order of purchases, the model achieves higher NDCG@5/10 than sequential models and GNNs. Our approach benefits from not needing to learn patterns from the noisy and heterogeneous complete purchase sequences, and it explicitly captures directed item-item relations, which are challenging for both sequential and graph models.

Additionally, the model is explainable and fast, making it competitive for real-world applications and an effective foundation for exploring more constrained approaches, such as personalizing the complementary item list and enhancing diversity. While we initially limited the complementary item list to those adjacent to the query item, we believe that leveraging global graph properties, like community transitions, could enhance recommendation performance.

There are, however, some relevant limitations to the model. It is built on the concept of product complementarity in e-commerce, where the growth rate of complementary items is much slower than the total number of items considered, limiting scalability in broader applications. Furthermore, the model is ID-based and does not project items into a latent space, which restricts its ability to predict complementary items for those not included in the training set. While we addressed this with semantic search, the cold-start problem remains a challenge.

\bibliography{main}
\bibliographystyle{IEEEtran.bst}

\end{document}